# Information Spreading in Dynamic Graphs

Andrea Clementi
Dipartimento di Ingegneria dell'Impresa
Università *Tor Vergata* di Roma
clementi@mat.uniroma2.it

Riccardo Silvestri
Dipartimento di Informatica
*Sapienza* Università di Roma
silvestri@di.uniroma1.it

Luca Trevisan
Department of Computer Science
Stanford University
trevisan@stanford.edu

October 22, 2018

**Abstract**

We present a general approach to study the flooding time (a measure of how fast information spreads) in dynamic graphs (graphs whose topology changes with time according to a random process). We consider arbitrary ergodic Markovian dynamic graph process, that is, processes in which the topology of the graph at time $t$ depends only on its topology at time $t-1$ and which have a unique stationary distribution. The most well studied models of dynamic graphs are all Markovian and ergodic.

Under general conditions, we bound the flooding time in terms of the mixing time of the dynamic graph process. We recover, as special cases of our result, bounds on the flooding time for the *random trip* model and the *random path* models; previous analysis techniques provided bounds only in restricted settings for such models. Our result also provides the first bound for the *random waypoint* model (which is tight for the most realistic ranges of network parameters) whose analysis had been an important open question.

**Keywords:** Dynamic Graphs, Flooding Protocols, Markov Chains, Mobility Models.

# 1 Introduction

A *dynamic graph* is a probabilistic process that describes a graph whose topology changes with time. Dynamic graphs are appropriate models of wireless networks, peer-to-peer networks, social networks, and so on. There are several interesting problems on dynamic graph processes, for example load balancing, studied in [16, 28]. Here, we are interested in the speed of *information spreading*, a question that can model the spread of disease, the broadcast of files on peer-to-peer networks, of memes in social networks, etc.

The simplest model of information spreading is the process of *flooding* [4, 10, 13, 15, 17, 18, 22, 26, 27], which begins with one node in the network being given a certain piece of information, and then the application of a protocol in which, at each time step, every node that has the information spreads it to its neighbors. Recall that the neighborhood of a node changes with time, and so even though the flooding algorithm is deterministic, the process of information spread is probabilistic.

The setting of wireless networks has motivated the study of this problem in geometric models of dynamic graphs in which, at every time step, every node is mapped to a point in a metric space, and two vertices are connected if their distance is smaller than a given communication radius. The underlying metric space is usually a bounded portion of the plane, for example a square or unit disk, and the dynamics come from independent random walks performed by the individual nodes via local moves [20, 15, 11, 12, 26, 27, 23]. For example, a representative model of this type is the random walk model: $n$ nodes are placed on a $m \times m$ grid; at each time step, every node $v$ independently moves to a point in the grid randomly chosen among the points adjacent to the one that $v$ occupied at the previous time step; at each time step, the edge $(u, v)$ is present in the dynamic graph if $u$ and $v$ are located within distance $r$ in the grid. In such a model, the flooding time will depend on the initial locations, on $r$, $m$ and $n$. Usually, one is interested in a worst-case analysis with respect to the initial locations, and in a bound dependent only on $r$, $m$ and $n$. Known analyses of such models rely rather strongly on the fact that, if we consider a fixed node $v$, the stationary distribution of its location (i.e. the positional distribution) on the grid under the random walk is essentially uniform.

The *random waypoint* model [7, 6, 8, 24] is another classic model of networks of mobile agents, and it is probably the most well studied one. In this model, every node chooses a random destination point in the mobility space, then he travels over the shortest path till he reaches the destination, and so on. Some analytical properties of this model have been derived such as the mixing time, the stationary node distribution, etc. [6, 8, 24]. However, bounding its flooding time (or any basic communication tasks such as routing, data collection, etc.) is still a fundamental open problem. The techniques adopted for the random walk model do not work in the random waypoint model, mainly because of the presence of long periods the node spends in deterministic trajectories and because of the strong difference between their respective stationary positional distributions. For instance, the stationary positional distribution of the random waypoint over a square is in fact far from the uniform one: it is highly biased towards the center of the square.

The above two families of models define a probabilistic process over the nodes, which then implies which pairs of nodes are connected by an edge. There are also models of probabilistic processes that are directly defined over the edges. A very general model is provided by Markovian Evolving Graphs (MEGs) introduced in [2] (see later for a formal definition). However, available techniques to analyze information spreading only concern very restricted subclasses of MEGs, such as that studied in [10], where the state (i.e. on/off) of the edges is ruled by independent copies of a simple two-state Markov chain.

## Our Work

- *General Dynamic Networks.* We provide an upper bound to the flooding time of any stationary Markovian Evolving Graph (MEG) [2]. If we call $G_t = (V, E_t)$ the random variable describing the dynamic graph at time $t$, the process is a MEG if the sequence of random variables $\mathcal{G}([n], \{E_t\}_{t \geqslant 0}) = G_0, \ldots, G_t, \ldots$ is Markovian, that is, if the distribution $G_t$ is completely determined by the distribution $G_{t-1}$, via a transformation independent of $t$. The class MEG is extremely general and, in particular, it includes all above-mentioned network models such as random walk and random waypoint: indeed, it is easy to verify that any of such models yields a sequence of random graphs $\mathcal{G}([n], \{E_t\}_{t \geqslant 0})$ which is Markovian. We require the Markovian sequence to converge to a unique stationary distribution, for every initial choice $G_0$. This stationary distribution is a probability distribution over $n$-vertex graphs and it will be called the stationary graph $\mathcal{G}_n$.

Our upper bound to the flooding time is a function of: i) the edge-probability in $\mathcal{G}_n$ (that determines the expected density of the stationary graph) ii) the degree of independence among edges in $\mathcal{G}_n$ and, iii) the mixing time of the Markov chain $\mathcal{G}([n], \{E_t\}_{t \geqslant 0})$. More precisely, given edge $e$, node $i$, and node subset $A$, let $p(e)$ be the probability that $e$ exists in $\mathcal{G}_n$ and let $e(i, A)$ be the binary random variable returning 1 iff an edge exists connecting $i$ to some node in $A$ in $\mathcal{G}_n$. Let $M$ be an upper bound on the mixing time of $\mathcal{G}([n], \{E_t\}_{t \geqslant 0})$. Our result states that if, for some arbitrary positive reals $\alpha$ and $\beta$, (i) $p(e) \geqslant \alpha$ for any link $e$ and[1] (ii) $\mathbf{P}\left(e(i, A) \cdot e(j, A)\right) \leqslant \beta \mathbf{P}\left(e(i, A)\right) \cdot \mathbf{P}\left(e(j, A)\right)$ for arbitrary nodes $i, j$ and arbitrary node subset $A$ (not containing $i$ or $j$), then the flooding time is w.h.p.[2]

$$O\left(M\left(\frac{1}{n\alpha} + \beta\right)^2 \log^2 n\right) \tag{1}$$

Our methods can be applied to non-Markovian processes as well, although the results are more complex to state (see Sections 2, 3). Bounding the mixing time of dynamic networks has been the subject of several studies in the last years [1, 24, 10]: our result allows to efficiently exploit any (previous or future) bound on the mixing time for the flooding time of such MEGs. In order to get an intuition of the real meaning of Conditions (i) and (ii) (i.e. how mild they can be), we observe that mild bounds on the density and independence parameters, say $\alpha = \Omega(1/n)$ and $\beta = O(\text{polylog} n)$, do not imply any good node/edge expansion of the single snapshot graphs $G_t$ of the process: In every $G_t$ there could be a large subset of all nodes (say half of them) that are isolated. In such sparse and disconnected topologies, thanks to our bound, the flooding time can be just a poly-logarithmic factor away from the mixing time of the MEG. This crucial fact has strong consequences on concrete network scenarios which are discussed below.

[1] With an abuse of notation, event probabilities such as $\mathbf{P}\left(e(i, A) = 1\right)$ will be shortly denoted as $\mathbf{P}\left(e(i, A)\right)$.

[2] We say that an event holds with high probability if it holds with probability at least $1 - 1/n$.

- *Node-MEGs.* Our general approach finds a natural application in a subclass of MEGs that we call node-Markovian Evolving Graphs, denoted as node-MEGs, where every node changes its state independently according to a Markov Chain $\mathcal{M}$. The state of a node can implement several dynamic features of the node (such as geometric position and destination, trajectory phase, social role, etc). Then the existence of a link between two nodes (only) depends on the current states of the two nodes according to a fixed deterministic function. Observe that node-MEGs are a relevant class of MEGs that includes every mobility model where nodes acts independently over any discrete space (such as an arbitrary graph or any geometric region discretized, for instance, by using a grid of suitable resolution[3]). Random walk, random waypoint, and random trip models [7, 24, 14] have a natural realization as node-MEGs: for instance, a realization of the random waypoint as a node-MEGs is described in Subsection 4.1. Notice that in node-MEGs, nodes are indistinguishable so Condition (i) is easy-to-check: if it is satisfied (in the stationary graph) for a specific edge then it is satisfied for all edges. We then prove that if incident edges are almost pairwise independent in the stationary graph yielded by a node-MEG then Condition (ii) is satisfied for some constant $\beta$: so, checking Condition (ii) can here be reduced to check pairwise independence of incident edges in a generic node.

- *Geometric Mobility Models.* As for node-MEGs defined over geometric spaces (such as the random waypoint), we state an easy-to-check condition implying Condition (ii): we transform the pairwise-independence condition on incident edges into some mild uniformity conditions on the *single-node* positional stationary distribution (see Corollary 4). We in fact show the former properties are satisfied by a wide class of random mobility models such as the random waypoint over a square: we thus get the first known upper bound for its flooding time. Furthermore, when the stationary graph is sparse, the obtained bound is almost tight, i.e., $O((L/v)\text{polylog} n)$ where $L$ is the diameter of the square and $v$ is the node speed. In particular, our bound is almost tight whenever the transmission radius and the node speed are absolute constants: this is surely the model setting that best fits opportunistic delay-tolerant Mobile Ad-hoc Networks [18, 19, 27].

- *Graph Mobility Models.* Our upper bound holds even when the mobility space is an arbitrary graph (the vertices of such graph are called points): in this case nodes choose randomly their trajectories from some fixed families of simple paths of the mobility graph and the mixing time is proportional to its hop-diameter. This mobility model is called *random paths* (on graphs). The parameter $\beta$ in Eq. 1 is here determined by the point congestion yielded by the feasible paths of the mobility graph: informally speaking, $\beta$ is small whenever the random paths do not yield a high point congestion (again, we get a somewhat mild uniformity condition on the (single-node) positional stationary distribution). If this is the case, the obtained bound on the flooding time becomes almost tight. The random walk model on graphs can be seen as a special case of the random paths on graphs: then the obtained bound on the flooding time improves over the previous one [15] for the class of graphs where the mixing time of a random walk is shorter than the meeting time of two random walks.

[3] The level of resolution does not affect the obtained bound on the flooding time, provided the resolution is high enough.

By concluding about mobile networks, we want to emphasize the impact of our general method over classic models: the general conditions $(i)$ and $(ii)$ of our bound are transformed into mild uniformity conditions over the positional stationary distribution which can be verified by standard techniques [1, 24].

- *Link-based Dynamic Networks.* In appendix we show that our method can also be applied to a subclass of MEGs in which each edge evolves independently according to an arbitrary (hidden) Markov chain. Previously [10, 5], such *link-based* dynamic models had been studied only in the case in which the edge Markov chain is very simple.

**Previous Works.** As mentioned before, information spreading in dynamic networks has been extensively studied in the literature under a variety of scenarios and objectives (for a recent good survey see [22]). For brevity's sake, we restrict our attention to the results more directly related to our work. Previous models and results can be roughly classified in two main classes: link-based dynamic graphs and mobility models.

As for the first class, in [9], radio broadcasting is analyzed on a dynamic graph managed by a worst-case dynamic adversary and on a sequence of independent Erdös-Rény graphs. In [10], an upper bound on the flooding time for the restricted model edge-MEG has been derived. The flooding time of another simple version of edge-MEGs has been studied in [5]. The general MEG model has been introduced in [2] where some results are obtained for the cover time and hitting time of random walks. Flooding time in stationary MEGs is studied in [11]; unlike our method, the method in [11] only holds for stationary graphs which are connected and good expanders. A worst-case model of dynamic graphs has been introduced in [21]. The analysis of some communication tasks is presented under the strong stability condition called $T$-interval connectivity (for $T \geqslant 1$) which stipulates that for every $T$ consecutive steps a stable connected spanning subgraph must exist.

As for mobility models, almost tight bounds on the flooding time for the random walk model have been obtained in [20, 11, 12, 26, 27, 23]. As above discussed, their techniques strongly rely on specific properties of the adopted version of the random walk. The case of general mobility graphs has been considered in [15]: the obtained results are discussed and compared to our results in Subsection 4.1. An upper bound on a variant of the random waypoint model has been derived in [13]. In this version, nodes follows *Manhattan* paths. Similarly to the works on the random walk models, the ad-hoc analysis in [13] strongly relies on the particular node trajectories and on the specific positional distribution yielded by this model. So, its contribution strongly departs from our general approach that obtains bounds for any version of the random waypoint model.

**Paper Organization.** Section 2 formalizes the general model of dynamic graph and the flooding process. Section 3 provides the main theorem for the flooding time in the general model. The node-MEG model is described in Section 4 where an upper bound on the flooding time is given for this specific model. The representation of the random trip and the random paths models as specific instances of the node-MEG model is given in Subsection 4.1: here, the flooding-time bound for node-MEGs is transformed into two useful bounds on the flooding time: the first one for the random trip and the second one for the random paths. Finally, conclusions with open questions are discussed in Section 5. Due to lack of space, the application of the main theorem to general edge-MEGs and all the proofs are given in the appendix.

## 2 Preliminaries

In this section, we introduce the general model of dynamic graphs. For any positive $n$, $[n]$ will denote the set $\{1, 2, \ldots, n\}$. A *dynamic graph* $\mathcal{G}([n], \{E_t\}_{t \geqslant 0})$, with node set $[n]$, is a stochastic process represented by a sequence of random variables $E_0, E_1, \ldots, E_t, \ldots$ such that, for every $t$, $E_t$ is the set of edges of the dynamic graph at time $t$. The speed of information spreading can be studied in terms of the flooding time. Given a dynamic graph $\mathcal{G}([n], \{E_t\}_{t \geqslant 0})$ and a node $s \in [n]$, the flooding process with source $s$ is defined as follows. At time $t = 0$, $s$ is the only informed node; then a node $v$ gets informed at time $t + 1$ iff an edge $e \in E_t$ exists connecting some informed node to $v$. Flooding over a dynamic graph is represented by the stochastic process $\{I_t\}_{t \geqslant 0}$ where

$$I_0 = \{s\} \quad \text{and} \quad \forall t \geqslant 1 \ I_t = I_{t-1} \cup \{j \in [n] \mid \exists i \in I_{t-1} : \{i, j\} \in E_t\}$$

The random variable $I_t$ is the set of informed nodes at time $t$. Clearly, it holds that

$$I_0 \subseteq I_1 \subseteq I_2 \subseteq \cdots I_t \subseteq \cdots$$

The *flooding time with source* $s$ is the random variable $F(\mathcal{G}, s) = \min_t\{I_t = [n]\}$ and the *flooding time* is the random variable $F(\mathcal{G}) = \max_s F(\mathcal{G}, s)$.
Given a dynamic graph $\mathcal{G}([n], \{E_t\}_{t \geqslant 0})$, we define the following random variables. For every time $t$, for every pair of nodes $i, j \in [n]$ and for every subset of nodes $A \subseteq [n]$, let

$$e_{i,j}^t = \begin{cases} 1 & \text{if } \{i, j\} \in E_t \\ 0 & \text{otherwise} \end{cases} \quad \text{and} \quad e_{i,A}^t = \begin{cases} 1 & \text{if } \exists j \in A : \{i, j\} \in E_t \\ 0 & \text{otherwise} \end{cases}$$

Moreover, for any binary random variable $X$, the notation $\mathbf{P}\left(X \mid E_{t \leqslant T}\right) \leqslant (\text{or} \geqslant)\ \alpha$ stands for

$$\forall \text{ sequence of edge subsets } A_0, \ldots, A_T, \quad \text{with } \mathbf{P}\left(\bigwedge_{t=0}^{T} (E_t = A_t)\right) > 0), \quad \text{it holds}$$

$$\mathbf{P}\left(X = 1 \;\middle|\; \bigwedge_{t=0}^{T} (E_t = A_t)\right) \leqslant (\text{or} \geqslant)\ \alpha$$

## 3 Flooding in Dynamic Graphs

Our goal is to evaluate the flooding time of a dynamic graph as a function of some properties of its edges. These properties are not required to show up at every snapshot, rather it suffices that they hold at the beginning of every time "epoch", where an epoch is a sequence of consecutive time steps of suitable length. When the dynamic graph is a Markovian process admitting a stationary graph, the properties above refer to the expansion properties of the stationary graph and the epoch length refers to the mixing time. However, aiming at the maximal generality, we introduce such concepts for general (non-Markovian) process.

Let $M$ be a positive integer and let $\alpha$ and $\beta$ be two positive reals. A dynamic graph $\mathcal{G}([n], \{E_t\}_{t \geqslant 0})$ is *($M$,$\alpha$,$\beta$)-stationary* if $\forall \tau \geqslant 1$, $\forall i, j \in [n]$ with $i \neq j$, $\forall A \subseteq [n] - \{i, j\}$, the following two conditions hold

1. $\mathbf{P}\left(e_{i,j}^{\tau M} \mid E_{t \leqslant (\tau-1)M}\right) \geqslant \alpha$ (Density Condition)
2. $\mathbf{P}\left(e_{i,A}^{\tau M} \cdot e_{j,A}^{\tau M} \mid E_{t \leqslant (\tau-1)M}\right) \leqslant \beta \mathbf{P}\left(e_{i,A}^{\tau M} \mid E_{t \leqslant (\tau-1)M}\right) \mathbf{P}\left(e_{j,A}^{\tau M} \mid E_{t \leqslant (\tau-1)M}\right)$ ($\beta$-Independence Condition)

The full proof of the following theorem is given in the Appendix.

**Theorem 1 (Flooding Time)** *If $\mathcal{G}([n], \{E_t\}_{t \geqslant 0})$ is $(M, \alpha, \beta)$-stationary then with high probability the flooding time in $\mathcal{G}$ is*

$$O\left(M\left(\frac{1}{n\alpha} + \beta\right)^2 \log^2 n\right)$$

# 4 Node Markovian Evolving Graphs

We introduce the general class of Node Markovian Evolving Graphs (in short, node-MEGs) where the behavior of the nodes is ruled by independent Markov chains. To every node is associated a Markov chain whose states contain enough information to determine whether two nodes are connected or not. Of course, this is an approach based upon hidden Markov chains to model dynamic graphs.

Let $\mathcal{M} = (S, P)$ be the Markov chain associated to every node, where $S$ is the set of states and $P : S \times S \to \mathbb{R}$ are the transition probabilities. The connections are determined by a symmetric map $C : S \times S \to \{0, 1\}$: any two nodes $i$, $j$ are connected at a given time $t$ if $C(u, v) = 1$, where $u$, $v$ are the states of $i$ and $j$ at time $t$. The symmetric map $C(\cdot, \cdot)$ is also called the connection graph of $\mathcal{M}$. A node-MEG is denoted by $NM(n, \mathcal{M}, C)$. Notice that the Markov chain $\mathcal{M}$ may depend on the number of nodes. The initial state of each node $i$ is random with a probability distribution $\iota_i$ over the set of states $S$. We denote by $\iota$ the global initial probability distribution determined by the product of the probability distributions $\iota_i$. The state of a node $i$ at time $t$ is a random variable $s_i^t$ fully determined by the initial distribution $\iota_i$ and the Markov chain $\mathcal{M}$. A node-MEG $NM(n, \mathcal{M}, C)$ together with an initial probability distribution $\iota$ determines a dynamic graph $\mathcal{G}([n], \{E_t\}_{t \geqslant 0})$ where, for any $t \geqslant 0$, $E_t = \{\{i, j\} \mid C(s_i^t, s_j^t) = 1\}$ It is easy to verify that node-MEGs enjoy the following property.

**Fact 2** *Consider any node-MEG* $\mathrm{NM} = NM(n, \mathcal{M}, C)$ *in its stationary state, then the probability* $P_{\mathrm{NM}}$ *that any fixed pair of nodes are connected and the probability* $P_{\mathrm{NM2}}$ *that two fixed nodes are both connected to another fixed one do not depend on the choice of the fixed nodes: they are functions only of the stationary distribution of* $\mathcal{M}$ *and of the symmetric map* $C(\cdot, \cdot)$.

In Subsection 4.1, we will show that a wide class of mobility models turns out to be a special instance of Node-MEGs.

**Flooding in Node-MEGs.** We now derive some simple properties ensuring that a node-MEG $NM(n, \mathcal{M}, C)$ is a $(M, \alpha, \beta)$-stationary dynamic graph. Since in a node-MEG, edges are not independent, the crucial condition is the $\beta$-independence. The models at hand are Markovian so the idea is to consider the model during its stationary state, that is, the time $M$ between two consecutive epochs is proportional to the mixing time of the Markov chain $\mathcal{M}$. Moreover, the $\beta$-independence involves sets of incident edges of arbitrary size (i.e. the size of subset $A$): instead, thanks to the independence of node evolutions and Fact 2, in node-MEGs this condition can be relaxed to a parameterized *pairwise* independence among incident edges.

**Theorem 3** *Let* $\mathrm{NM} = NM(n, \mathcal{M}, C)$ *be a node-MEG such that* $P_{\mathrm{NM}} \geqslant 1/n^{\mathrm{O}(1)}$ *and* $P_{\mathrm{NM}2} \leqslant \eta(P_{\mathrm{NM}})^2$*, for some* $\eta \geqslant 1$*. Then, with high probability, the flooding time is*

$$O\left(T_{\mathrm{mix}}\left(\frac{1}{nP_{\mathrm{NM}}} + \eta\right)^2 \log^3 n\right)$$

*where* $T_{\mathrm{mix}}$ *is the mixing time of the Markov chain* $\mathcal{M}$*.*

## 4.1 Flooding in Classic Mobility Models

**Geometric Mobility Models.** Several mobility models can be represented as special cases of node-MEGs. Many of these are continuous-space models [7, 24] in which nodes move over a subset of $\mathbb{R}^d$. Since node-MEGs are discrete, we approximate continuous space by discretization. In the simplest and most common case, nodes move over a square of $\mathbb{R}^2$ of side length $L$. The square can be discretized by taking a square grid $\mathcal{Q}$ formed by $m \times m$ points regularly spaced in the square region, where $m$ can be arbitrarily chosen.

In the standard random waypoint [7], $n$ nodes independently move over the square: every node randomly chooses a speed in $[v_{min}, v_{max}]$ where $v_{max} = \Theta(v_{min})$ and a destination point ('waypoint') in the square and moves with the chosen speed on a straight path to this point. Then, it repeats the same process again and again. The destination points are uniformly distributed on the square. At any time two nodes are connected if they are at distance not larger than the transmission radius $r$.

The formal discretization of the random waypoint as a node-MEG can be done by simple and standard arguments [14], so we here provide an informal description only. The generic state of the Markov chain $\mathcal{M}$ must encode the destination point, the current point in the straight point-path the node lies, and the node speed (the latter can be defined as the number of points per time step). Then the transition matrix can be easily defined: when a node is in some internal point of a path the choice of his next state is deterministic while when he arrives at the end of a path, his next state is randomly chosen by selecting the next destination point (and thus the next path to be followed) and the speed. As usual, the connection map $C$ is defined as follows: there is an edge between nodes $u$ and $v$ at time $t$ iff their relative distance at time $t$ is not larger than $r$. The *positional* probability distribution in the stationary phase is defined as the probability that a node is in point $x$ (for any choice of $x$ over the square) when the state of the node is random with the stationary distribution of $\mathcal{M}$. The density function of this distribution (in short, *positional function*) yielded by the random waypoint over the square will be denoted by $F_{wp}(\cdot)$. The random waypoint belongs to a general class of geometric mobility models called the *random trip model* [24] where the mobility space $\mathcal{R}$ can be any bounded connected region of $\mathbb{R}^d$ and the feasible node-trajectories can be any family of continuous curves. Any random trip model can be discretized with any level of "approximation" (in terms of grid resolution and time unit) by following the same procedure described above for the random waypoint. For this geometric class of node-MEGs, Theorem 3 can be rewritten in a very useful way.
For any $r \geqslant 0$, $D(u,r)$ denotes the set of all the points that are at Euclidean distance at most $r$ from $u$. For any connected region $\mathcal{B} \subseteq \mathbb{R}^d$, define $\mathcal{B}_r = \{u \in \mathcal{B} \mid D(u,r) \subseteq \mathcal{B}\}$ and $\mathrm{vol}(\mathcal{B})$ as the volume of region $\mathcal{B}$. The proof of the next bound is given in the Appendix.

**Corollary 4** *Let* $\mathrm{NM} = NM(n, \mathcal{M}, C)$ *be a node-MEG yielded by a suitable discretization of a random trip model* $\mathcal{T}$ *over a bounded connected region* $\mathcal{R} \subseteq \mathbb{R}^d$ *with positional function* $F_{\mathcal{T}}$*. If* $P_{\mathrm{NM}} \geqslant 1/n^{\mathrm{O}(1)}$ *and for some* $\delta \geqslant 1$ *and* $\lambda > 0$ *it holds that*

*(a)* $\forall u \in \mathcal{R}$, $F_{\mathcal{T}}(u) \leqslant \frac{\delta}{\mathrm{vol}(\mathcal{R})}$

*(b)* $\exists \mathcal{B} \subseteq \mathcal{R}$ *such that* $\mathrm{vol}(\mathcal{B}_r) \geqslant \lambda \mathrm{vol}(\mathcal{R})$ *and* $\forall u \in \mathcal{B}$, $F_{\mathcal{T}}(u) \geqslant \frac{1}{\delta \mathrm{vol}(\mathcal{R})}$

*then, with high probability, the flooding time is* $\mathrm{O}\left(T_{\mathrm{mix}} \left(\frac{\delta^2 \mathrm{vol}(\mathcal{R})}{\lambda n r^d} + \frac{\delta^6}{\lambda^2}\right)^2 \log^3 n\right)$ *where* $T_{\mathrm{mix}}$ *is the mixing time of the Markov chain* $\mathcal{M}$*.*

The useful novelty of the above corollary lies in the following fact: the pairwise-independence condition in Theorem 3 is transformed into two mild "uniformity" conditions on the positional function yielded by the mobility model. The latters only refer to the stationary positional distribution of the single node and it is often much easier to verify with respect to the pairwise condition. Indeed, a general method (the Palm Calculus) to derive explicit formulas of such function for random trip models has been introduced in [24]. As for the random waypoint on the square, the explicit positional function $F_{wp}(\cdot)$ has been derived in [25] and it is easy to verify that the two conditions of the above corollary are satisfied for some absolute constants $\delta$ and $\lambda$. Furthermore, the mixing time of the random waypoint over a square of side length $L$ is $\Theta(L/v_{max})$ (remind we are assuming $v_{max} = O(v_{min})$) [1, 29]. We thus obtain the following bound on the flooding time $O\left(\frac{L}{v_{max}} \left(\frac{L^2}{nr^2} + 1\right)^2 \log^3 n\right)$ Let us consider the case $L \sim \sqrt{n}$, $r = \Omega(1)$, and $r = O(v_{max})$; notice that this standard setting yields a stationary mobile network which is w.h.p. sparse and highly disconnected. Then the bound on the flooding time becomes $O\left(\frac{\sqrt{n}}{v_{max}} \log^3 n\right)$ which almost matches the trivial lower bound $\Omega\left(\frac{\sqrt{n}}{v_{max}}\right)$.

**Graph Mobility Models.** A natural generalization of random walks over a graph can be defined by considering random paths over a graph. This clearly includes the random waypoint over a graph. At every time step, a node moves along a path instead of on a single edge. More precisely, the model is specified by a graph $H(V, A)$ and a family $\mathcal{P}$ of feasible paths in $H$ satisfying the property: for every path $h \in \mathcal{P}$, there is a path $h' \in \mathcal{P}$ such that $h'$ starts where $h$ ends. For any $u \in V$, let $\mathcal{P}(u)$ be the set of paths in $\mathcal{P}$ that starts at point $u$. The mobility model is as follows, a node at point $u \in V$ chooses uniformly at random a path in $\mathcal{P}(u)$, then it travels along the path (an edge at the time), when it reaches the end point $v$, it chooses uniformly at random a path in $\mathcal{P}(v)$ and travels along that path, and so on. We assume that two nodes are connected, at any given time $t$, if they are in the same point at time $t$. For any path $h$, let $\ell(h)$ denote the number of points of $h$. The representation of a random path model RP $= (H, \mathcal{P})$ as a node-MEG is straightforward. The Markov chain $\mathcal{M}_{\mathrm{RP}} = (S, P)$ is such that $S = \{(h, h_i) \mid h \in \mathcal{P},\ 2 \leqslant i \leqslant \ell(h)$ and $h_i$ is the $i$th point of $h\}$; the transition probabilities are as follows

$$\begin{aligned}
&\forall h \in \mathcal{P}\ \forall i : 2 \leqslant i < \ell(h) \qquad P((h, h_i), (h, h_{i+1})) = 1 \\
&\forall h, h' \in \mathcal{P} : h_{\ell(h)} = h'_1 \qquad P((h, h_{\ell(h)}), (h', h'_2)) = \frac{1}{|\mathcal{P}(h_{\ell(h)})|} \\
&\text{all other transition probabilities are equal to zero;}
\end{aligned}$$

and the connection map is such that $C_{\mathrm{RP}}((h, h_i), (h', h'_j)) = 1$ iff $h_i = h'_j$. Observe that if $\mathcal{P}$ is the set of edges of $H$ then the mobility model is equivalent to the random walk over $H$ (with $\rho = 1$).

We say that a path $h \in \mathcal{P}$ *passes through* a point $u$ if $h_i = u$ for some $2 \leqslant i \leqslant \ell(h)$. For any point $u \in V$, let $\#_{\mathcal{P}}(u)$ be the number of paths in $\mathcal{P}$ that passes through point $u$. Notice that if $\mathcal{P}$ is the set of edges of $H$, then $\#_{\mathcal{P}}(u) = \deg_H(u)$. A random-path model RP $= (H, \mathcal{P})$ is said to be *simple* if every path in $\mathcal{P}$ does not pass through the same point more than once, but the start and end points that may be equals. Moreover, RP is *reversible* if, for every path $h \in \mathcal{P}$, the reverse path of $h$ also belongs to $\mathcal{P}$. We say that a random path model $(H, \mathcal{P})$ is *$\delta$-regular* if

$$\forall u \in V \qquad \#_{\mathcal{P}}(u) \leqslant \delta \frac{\sum_{v \in V} \#_{\mathcal{P}}(v)}{|V|}$$

Roughly speaking $\delta$-regularity ensures that no point is a much busier crossroad than the average. Theorem 3 then implies the following useful result (the proof is given in the Appendix).

**Corollary 5** *Let* NM $= NM(n, \mathcal{M}, C)$ *be a node-MEG yielded by a random path model* RP $=$ $(H, \mathcal{P})$ *that is simple, reversible, $\delta$-regular, and $|V| \leqslant n^{\mathrm{O}(1)}$. Then, w.h.p the flooding time is* $\mathrm{O}\left(T_{\mathrm{mix}} \left(\frac{|V|}{n} + \delta^3\right)^2 \log^3 n\right)$ *where $T_{\mathrm{mix}}$ is the mixing time of the Markov chain $\mathcal{M}$.*

If for every pair of points there is only one feasible simple path, then the mixing time of the relative Markov chain is $\mathrm{O}(D)$, where $D$ is the diameter of $H$. Moreover, if the model is $\delta$-regular for some $\delta = \mathrm{polylog}(n)$ and $|V| = \mathrm{O}(n\mathrm{polylog}(n))$, then the above corollary implies that the flooding time is $O(D\ \mathrm{polylog}(n))$. This is within a poly-logarithmic factor from the optimum, since a trivial lower bound is $\Omega(D)$. A basic instance of this case is when $H$ is a grid and the feasible paths are the shortest ones.

As mentioned in the Introduction, almost tight bounds for the flooding time on the random walk model over grids have been recently obtained in [11, 12, 27]. In a general graph $H(V, A)$, every node randomly chooses his next position among all points in $V$ that are within $\rho$ hops from his current position. The transmission radius $r$ determines the maximal distance (again in terms of number of hops in $H(V, A)$) within which a message can be successfully transmitted. The most studied setting is $\rho = 1$ and $r = 0$: a node makes at most one hop per time step and it can infect only nodes that lies in the same point. This natural setting in general graphs has been studied in [15]: the flooding time is proved to be $O(T^* \log n)$ where $T^*$ is the *meeting* time between two independent random walks on $H$.

In what follows we apply our analysis for the random path model to the special case of random walks. The $\delta$-regularity condition over paths is transformed into a simple condition on the degree of the points. Given any $\delta \geqslant 1$, a graph $H(V, A)$ is said to be $\delta$-regular if $(\max\{\deg(v) \mid v \in V\})/(\min\{\deg(v) \mid v \in V\}) \leqslant \delta$. Then, we can derive a simple adaptation of Corollary 5.

**Corollary 6** *Let* $\mathrm{NM} = NM(n, \mathcal{M}, C)$ *be a node-MEG yielded by the random walk over any* $\delta$*-regular mobility graph* $H(V, A)$*. Then, w.h.p the flooding time is* $O\left(T_{\mathrm{mix}}\left(\frac{\delta^2|V|}{n} + \delta^7\right)^2 \log^3 n\right)$ *where* $T_{\mathrm{mix}}$ *is the mixing time of a random walk over* $H$*.*

The above bound improves the result in [15] over a relevant and wide class of mobility graphs. Indeed, given a symmetric graph, the meeting time of two random walks is asymptotically equivalent to the first hitting time and can be much larger than the mixing time of a single random walk [1]. A natural example is that of $k$-augmented grids: take a grid of $s$ points and add an edge between any pair of points whose hop-distance is not larger than $k$. While the meeting time is not smaller than that of a standard grid $\Theta(s \log s)$ [1, 27], the mixing time decreases in $k$. For instance, if $s \sim n\mathrm{polylog}(n)$ then the bound in [15] becomes $O(n\mathrm{polylog}(n))$ while our bound is $O((n\mathrm{polylog}(n))/k^2)$.

# 5 Conclusions

We believe that significant improvements are possible to the bound in Theorem 1 along some interesting directions. We suspect that, under mild assumptions on the dynamic graph process, the factor $(\frac{1}{n\alpha} + \beta)^2$ can be improved. A more challenging task is to avoid the dependency of the bound on the mixing time of the graph process. The density and $\beta$-independence conditions can be met even at a state in which the graph process is far from the stationary distribution, and so a more refined analysis might be able to bound the flooding time without having to"wait" for the process to achieve stationarity.
Our method may prove useful in the analysis of more refined communication protocols than flooding. A simple example is a randomized protocol in which, at every step, a node that possesses the information transmits it to a randomly chosen subset of neighbors. The analysis of such a process can be reduced to the analysis of flooding in a "virtual" dynamic graph in which a subset of the edges are removed. More general communication protocols might also be reduced to flooding by folding the actions of the protocol into the dynamic graph process.

# A Generalized Edge-MEGs

The link-based dynamic model Edge-MEG has been introduced in [10] and successively studied in [4, 11, 17]. In this restricted instance of MEGs, edges evolves independently. Every edge of the $n$-node graph can be in two states only: *on* or *off.* At any time step, every edge changes its state according to a two-state Markovian process with probabilities $p$ (edge birth-rate) and $q$ (edge death-rate): if at time $t$ an edge $e$ exists then it will die with probability $q$ while if $e$ does not exists then it will come up with proability $p$. In [10], the authors prove an almost tight upper bound on the flooding time.

$$O(\log n / \log(1 + np)) \tag{2}$$

A more refined model with four states has been recently introduced and studied in [5]. Edge-MEGs do not explicitly model node's mobility, rather they are more suitable to model the link evolution in peer-to-peer networks or faulty networks.

Our main contribution here lies in the fact that Theorem 1 can be applied to the much more general version of edge-MEGs where an arbitrary (hidden) Markov chain $\mathcal{M} = (S, P)$ rules the behavior of every edge and by an arbitrary map $\chi : S \to \{0, 1\}$ that determines, in function of the state, whether the edge exists or not. This wide generalization of edge-MEG will be denoted as $EM(n, \mathcal{M}, \chi)$.

The initial state of each edge $\{i, j\}$ is random with a probability distribution $\iota_{\{i,j\}}$ over the set of states $S$. We denote by $\iota$ the global initial probability distribution determined by the product of the probability distributions $\iota_{\{i,j\}}$. The state of an edge $\{i, j\}$ at any time $t$ is a random variable $s^t_{\{i,j\}}$ completely determined by the initial distribution $\iota_{\{i,j\}}$ and the Markov chain $\mathcal{M}$.

Any model $EM(n, \mathcal{M}, \chi)$ together with an initial probability distribution $\iota$ determines a dynamic graph $\mathcal{G}([n], \{E_t\}_{t \geqslant 0})$ where, for any $t \geqslant 0$,

$$E_t = \{\{i, j\} \mid \chi(s^t_{\{i,j\}}) = 1\}$$

A crucial property of such generalized edge-MEGs is that edges are independent random variables, so it always holds that the $\beta$-indpendence is satisfied with $\beta = 1$. Then, when the Markov chain $\mathcal{M}$ admits a unique stationary distribution $\pi$, Theorem 1 implies that flooding time is

$$O\left(T_{\text{mix}} \left(\frac{1}{n\alpha} + 1\right)^2 \log^2 n\right)$$

where $T_{\text{mix}}$ is the mixing time of the Markov chain $\mathcal{M}$ and $\alpha$ is the probability an edge exists in the stationary regime (i.e. according to $\pi$). For instance, in the basic edge-MEG model with parameters $p$ and $q$ the mixing time is $\Theta(1/(p + q))$ and $\alpha = p/(p + q)$ [10]. We thus get an upper bound

$$O\left(\frac{1}{p+q} \left(\frac{p+q}{np} + 1\right)^2 \log^2 n\right)$$

By comparing our bound to the (almost-tight) one in Eq. 2, we get that the former is almost tight whenever $q \geqslant np$.

## B Some useful Inequalities

**The Paley-Zigmund inequality.** If $X \geqslant 0$ is a random variable with finite variance, and if $0 < \theta < 1$, then

$$\mathbf{P}\left(X \geqslant \theta \mathbf{E}\left[X\right]\right) \;\geqslant\; (1-\theta^2)\frac{\mathbf{E}\left[X\right]^2}{\mathbf{E}\left[X^2\right]} \tag{3}$$

**Lemma 7 (Lemma 3.1 in [3])** *Let $X_1, \ldots, X_n$ be a sequence of random variables with values in an arbitrary domain, and let $Y_1, \ldots, Y_n$ be a sequence of binary random variables, with the property that $Y_i = Y_i(X_1, \ldots, X_n)$. If*

$$\mathbf{P}\left(Y_i = 1 \mid X_1, \ldots, X_{i-1}\right) \geqslant p$$

*then*

$$\mathbf{P}\left(\sum Y_i \leqslant k\right) \;\leqslant\; \mathbf{P}\left(B(n,p) \leqslant k\right)$$

*where $B(n,p)$ is binomially distributed random variable with parameters $n$ and $p$.*

**Lemma 8 (Chernoff Bound)** *Let $X_1, \ldots X_n$ be independent binary random variables and let $X = \sum_{i=1}^{n} X_i$ and $\mu = \mathbf{E}\left[X\right]$. Then, for any $\delta > 0$,*

$$\mathbf{P}\left(X < (1-\delta)\mu\right) \;<\; \exp\left(-\frac{\delta^2 \mu}{2}\right)$$

## C Proof of Theorem 1

**Expansion Properties.** In what follows, we derive some expansion properties of $(M,\alpha,\beta)$-stationary dynamic graphs. Such properties will be then exploited in the analysis of flooding.
The times $\tau M$ will be called *epochs* and they will be abbreviated by $\tau$ (i.e. $\tau$ will stand for $\tau M$, with respect to a fixed $(M, \alpha, \beta)$-stationary dynamic graph $\mathcal{G}([n], \{E_t\}_{t \geqslant 0})$). Thus, we write $E_\tau$, $e_{i,j}^{\tau}$ and $e_{i,A}^{\tau}$ to denote $E_{\tau M}$, $e_{i,j}^{\tau M}$ and $e_{i,A}^{\tau M}$, respectively. These abbreviations will be also used for other random variables.
Let $\deg_{i,A}^{\tau}$ be the random variable counting the number of nodes in $A$ connected to $i$ at epoch $\tau$, i.e., $\deg_{i,A}^{\tau} = |\{j \in A \mid \{i,j\} \in E_\tau\}|$. Observe that, thanks to Condition (1), the expected value of $\deg_{i,A}^{\tau}$ is at least $|A|\alpha$; the following lemma provides a concentration result as function of the "independence" parameter $\beta$.

**Lemma 9** *If $\mathcal{G}([n], \{E_t\}_{t \geqslant 0})$ is $(M, \alpha, \beta)$-stationary then, $\forall \tau \geqslant 1$, $\forall i \in [n]$, and $\forall A \in [n] - \{i\}$,*

$$\mathbf{P}\left(\deg_{i,A}^{\tau} \geqslant \frac{|A|\alpha}{2} \;\middle|\; E_{\leqslant \tau - 1}\right) \;\geqslant\; \frac{|A|\alpha}{2 + 2|A|\alpha\beta}$$

*Proof.* Firstly we bound the expected square of the degree. For the sake of brevity, we omit the conditioning by $E_{\leqslant \tau-1}$. It holds that

$$
\begin{aligned}
\mathbf{E}\left[\left(\deg_{i,A}^{\tau}\right)^2\right] &= \mathbf{E}\left[\left(\sum_{j\in A} e_{i,j}^{\tau}\right)^2\right] \\
&= \sum_{j,k\in A} \mathbf{E}\left[e_{i,j}^{\tau}\cdot e_{i,k}^{\tau}\right] \\
&= \sum_{j,k\in A, j\neq k} \mathbf{E}\left[e_{i,j}^{\tau}\cdot e_{i,k}^{\tau}\right] + \mathbf{E}\left[\deg_{i,A}^{\tau}\right] \qquad (\text{since } (e_{i,j}^{\tau})^2 = e_{i,j}^{\tau}) \\
&\leqslant \sum_{j,k\in A, j\neq k} \beta\mathbf{E}\left[e_{i,j}^{\tau}\right]\mathbf{E}\left[e_{i,k}^{\tau}\right] + \mathbf{E}\left[\deg_{i,A}^{\tau}\right] \qquad (\text{by Condition (2)}) \\
&\leqslant \beta\mathbf{E}\left[\deg_{i,A}^{\tau}\right]^2 + \mathbf{E}\left[\deg_{i,A}^{\tau}\right]
\end{aligned}
$$

Moreover, from Condition (1), it derives that

$$
\mathbf{E}\left[\deg_{i,A}^{\tau}\right] \geqslant |A|\alpha
$$

Thus, from the Paley-Zigmund inequality (with $\theta = 1/2$) and the above bounds, we obtain

$$
\begin{aligned}
\mathbf{P}\left(\deg_{i,A}^{\tau} \geqslant \frac{|A|\alpha}{2}\right) &\geqslant \frac{1}{2}\frac{\mathbf{E}\left[\deg_{i,A}^{\tau}\right]^2}{\mathbf{E}\left[(\deg_{i,A}^{\tau})^2\right]} \\
&\geqslant \frac{1}{2}\frac{\mathbf{E}\left[\deg_{i,A}^{\tau}\right]^2}{\beta\mathbf{E}\left[\deg_{i,A}^{\tau}\right]^2 + \mathbf{E}\left[\deg_{i,A}^{\tau}\right]} \\
&= \frac{\mathbf{E}\left[\deg_{i,A}^{\tau}\right]}{2+2\beta\mathbf{E}\left[\deg_{i,A}^{\tau}\right]} \\
&\geqslant \frac{\mathbf{E}\left[\deg_{i,A}^{\tau}\right]}{2+2\beta\mathbf{E}\left[\deg_{i,A}^{\tau}\right]}
\end{aligned}
$$

Since function $\frac{x}{2+2\beta x}$ is decreasing, we have that

$$
\mathbf{P}\left(\deg_{i,A}^{\tau} \geqslant \frac{|A|\alpha}{2}\right) \geqslant \frac{\mathbf{E}\left[\deg_{i,A}^{\tau}\right]}{2+2\beta\mathbf{E}\left[\deg_{i,A}^{\tau}\right]} \geqslant \frac{|A|\alpha}{2+2\beta|A|\alpha}
$$

□

The next lemma extends Lemma 9 to the expansion of an arbitrary node subset. For every epoch $\tau$ and for every $A, B \subseteq [n]$, define the random variable counting the expansion from $A$ to $B$

$$
\deg_{A,B}^{\tau} = |\{j \in B \mid \exists i \in A : \{i,j\} \in E_{\tau}\}|
$$

**Lemma 10** *If $\mathcal{G}([n], \{E_t\}_{t\geqslant 0})$ is $(M,\alpha,\beta)$-stationary then, $\forall \tau \geqslant 1$, $\forall A \in [n]$, and $\forall B \in [n] - A$,*

$$
\mathbf{P}\left(\deg_{A,B}^{\tau} \geqslant \frac{|A||B|\alpha}{4+4|A|\alpha\beta} \;\middle|\; E_{\leqslant\tau-1}\right) \geqslant \frac{|A||B|\alpha}{4+6|A||B|\alpha\beta}
$$

*Proof.* The argument is very similar to that of the proof of Lemma 9. Firstly we bound the expected square of the expansion. For the sake of brevity, we omit the conditioning by $E_{\leqslant \tau-1}$.

$$\begin{aligned}
\mathbf{E}\left[(\deg^{\tau}_{A,B})^2\right] &= \mathbf{E}\left[\left(\sum_{j\in B} e^{\tau}_{j,A}\right)^2\right] \\
&= \sum_{j,k\in B} \mathbf{E}\left[e^{\tau}_{j,A}\cdot e^{\tau}_{k,A}\right] \\
&\leqslant \mathbf{E}\left[\deg^{\tau}_{A,B}\right] + \beta \sum_{j,k\in B} \mathbf{E}\left[e^{\tau}_{j,A}\right]\mathbf{E}\left[e^{\tau}_{k,A}\right] \quad \text{(by Condition (2))} \\
&= \mathbf{E}\left[\deg^{\tau}_{A,B}\right] + \beta \mathbf{E}\left[\deg^{\tau}_{A,B}\right]^2
\end{aligned}$$

Moreover, from Lemma 9 it derives that

$$\mathbf{E}\left[\deg^{\tau}_{A,B}\right] = \sum_{j\in B}\mathbf{E}\left[e^{\tau}_{j,A}\right] = \sum_{j\in B}\mathbf{P}\left(\deg^{\tau}_{j,A} > 0\right) \geqslant \sum_{j\in B}\frac{|A|\alpha}{2+2|A|\alpha\beta} = \frac{|A||B|\alpha}{2+2|A|\alpha\beta}$$

From the Paley-Zigmund inequality (with $\theta = 1/2$) and the above bounds, we have that

$$\begin{aligned}
\mathbf{P}\left(\deg^{\tau}_{A,B} \geqslant \frac{|A||B|\alpha}{4+4|A|\alpha\beta}\right) &\geqslant \frac{1}{2}\frac{\mathbf{E}\left[\deg^{\tau}_{A,B}\right]^2}{\mathbf{E}\left[(\deg^{\tau}_{A,B})^2\right]} \\
&\geqslant \frac{\mathbf{E}\left[\deg^{\tau}_{A,B}\right]^2}{2\mathbf{E}\left[\deg^{\tau}_{A,B}\right] + 2\beta\mathbf{E}\left[\deg^{\tau}_{A,B}\right]^2} \\
&= \frac{\mathbf{E}\left[\deg^{\tau}_{A,B}\right]}{2+2\beta\mathbf{E}\left[\deg^{\tau}_{A,B}\right]} \\
&\geqslant \frac{\frac{|A||B|\alpha}{2+2|A|\alpha\beta}}{2+2\beta\frac{|A||B|\alpha}{2+2|A|\alpha\beta}} \\
&\geqslant \frac{|A||B|\alpha}{4+6|A||B|\alpha\beta}
\end{aligned}$$

□

When the dynamic graph is sparse, the expansion rate obtained by considering a single snapshot of the process (i.e. the expansion of a node subset at time $\tau$) does not suffice to get a good number of new informed nodes. In this case, a "dynamic" version of the expansion properties is required. For every epoch $\tau$, for every $T \geqslant 1$, and for every $A \subseteq [n]$, define

$$\text{spread}^{\tau,T}_{A} = |\{j \in [n] - A \mid \exists \tau' \exists i \in A : \tau < \tau' \leqslant \tau + T \wedge \{i,j\} \in E_{\tau'}\}|$$

That is, $\text{spread}^{\tau,T}_{A}$ is the number of nodes outside $A$ that get connected to nodes in $A$ during the epochs in the interval $(\tau, \tau+T]$.

**Lemma 11** *If $\mathcal{G}([n], \{E_t\}_{t\geqslant 0})$ is $(M, \alpha, \beta)$-stationary then, $\forall \tau \geqslant 1$, $\forall A \subseteq [n]$ with $|A| \leqslant n/4$, and $\forall t \geqslant 0$,*

$$\mathbf{P}\left(\text{spread}_A^{\tau,T} < |A| \;\middle|\; E_{\leqslant\tau}\right) \;<\; \exp(-t)$$

*where*

$$T = 256\left(\frac{1}{|A|n^2\alpha^2} + \frac{\beta}{n\alpha} + \frac{|A|\beta^2}{n}\right) + \left(\frac{4}{|A|n\alpha} + 3\beta\right)t$$

*Proof.* For brevity's sake, we omit the conditioning by $E_{\leqslant\tau}$. Let $S_t$ be the set of nodes outside $A$ that get connected to nodes in $A$ during the epochs in $(\tau, \tau+t]$. Formally,

$$S_0 = \emptyset \quad \text{and} \quad S_t = S_{t-1} \cup \{j \in [n] - A \mid \exists i \in A : \{i,j\} \in E_{\tau+t}\}$$

Clearly, $|S_T| = \text{spread}_A^{\tau,T}$. Let

$$\gamma = \frac{|A|n\alpha}{8 + 8|A|\alpha\beta}$$

Define

$$Y_t = \begin{cases} 1 & \text{if } |S_{t-1}| \geqslant |A| \text{ or } |S_t| \geqslant |S_{t-1}| + \lceil\gamma\rceil \\ 0 & \text{otherwise} \end{cases}$$

Observe that $Y_t = f_t(E_{\tau+1}, \ldots, E_{\tau+t})$ for a suitable function $f_t$. From the inequality $\mathbf{P}\left(A \vee B \mid H\right) \geqslant \mathbf{P}\left(B \mid H \wedge \overline{A}\right)$ it derives that

$$\begin{aligned} \mathbf{P}\left(Y_t = 1 \mid E_{\tau,t-1}\right) &= \mathbf{P}\left(|S_{t-1}| \geqslant |A| \;\vee\; |S_t| > |S_{t-1}| + \lceil\gamma\rceil \mid E_{\tau,t-1}\right) \\ &\geqslant \mathbf{P}\left(|S_t| > |S_{t-1}| + \lceil\gamma\rceil \mid E_{\tau,t-1} \;\wedge\; |S_{t-1}| < |A|\right) \end{aligned} \quad (4)$$

where $E_{\tau,t-1}$ stands for $E_{\tau+1}, \ldots, E_{\tau+t-1}$. Assume that $|S_{t-1}| < |A|$ and let $W = [n]-A-S_{t-1}$. Since $|A| \leqslant n/4$, it holds that $|W| \geqslant \frac{n}{2}$ and

$$|S_t| > |S_{t-1}| + \lceil\gamma\rceil \quad \Leftrightarrow \quad \deg_{A,W}^{\tau+t} \geqslant \gamma \quad (5)$$

From Lemma 10, we have that

$$\begin{aligned} \mathbf{P}\left(\deg_{A,W}^{\tau+t} \geqslant \gamma \;\middle|\; E_{\leqslant\tau+t-1} \wedge |S_{t-1}| < |A|\right) &\geqslant \mathbf{P}\left(\deg_{A,W}^{\tau+t} \geqslant \frac{|A||W|\alpha}{4+4|A|\alpha\beta} \;\middle|\; E_{\leqslant\tau+t-1} \wedge |S_{t-1}| < |A|\right) \\ &\geqslant \frac{|A||W|\alpha}{4+6|A||W|\alpha\beta} \\ &\geqslant p = \frac{|A|n\alpha}{8+6|A|n\alpha\beta} \end{aligned} \quad (6)$$

Thus, from Ineq.s 4, 5, and 6 we get

$$\mathbf{P}\left(Y_t = 1 \mid E_{\tau,t-1}\right) \;\geqslant\; p$$

We can now apply Lemma 7 to the r.v. $E_{\tau+1}, \ldots, E_{\tau+T}$ and r.v. $Y_1, \ldots, Y_T$

$$\mathbf{P}\left(\sum_{t=1}^{T} Y_t < \frac{|A|}{\gamma}\right) \;\leqslant\; \mathbf{P}\left(B(T,p) < \frac{|A|}{\gamma}\right)$$

Since $\mathrm{spread}_A^{\tau,T} < |A| \;\Rightarrow\; \sum_{t=1}^T Y_t < \frac{|A|}{\gamma}$, from the above inequality we obtain

$$\mathbf{P}\left(\mathrm{spread}_A^{\tau,T} < |A|\right) \;\leqslant\; \mathbf{P}\left(B(T,p) < \frac{|A|}{\gamma}\right)$$

By applying Chernoff's Bound (Lemma 8), after some calculations, we get

$$\mathbf{P}\left(B(T,p) < \frac{|A|}{\gamma}\right) \;\leqslant\; \exp(-t) \;\text{ for }\; T = 256\left(\frac{1}{|A|n^2\alpha^2} + \frac{\beta}{n\alpha} + \frac{|A|\beta^2}{n}\right) + \left(\frac{4}{|A|n\alpha} + 3\beta\right)t$$

□

The next result still concerns the "dynamic" expansion of any subset of nodes. It will be applied when the subset of informed node is large (say at least $n/2$) in order to get a good bound on the completion time of the last phase of the flooding process. Let us define the following r.v.

$$e_{i,A}^{\tau,T} = \begin{cases} 1 & \text{if } \exists \tau' \; \exists j \in A \; : \tau < \tau' \leqslant \tau + T \;\text{ and }\; \{i,j\} \in E_{\tau'} \\ 0 & \text{otherwise} \end{cases} \tag{7}$$

**Lemma 12** *If $\mathcal{G}([n], \{E_t\}_{t\geqslant 0})$ is $(M,\alpha,\beta)$-stationary then, $\forall \tau, t \geqslant 1$ it holds that, for every $A \subseteq [n]$ and for every $i \in [n] \setminus A$, it holds that*

$$\mathbf{P}\left(e_{i,A}^{\tau,T} = 0 \;\middle|\; E_{\leqslant\tau}\right) \;\leqslant\; \exp(-t), \quad \textit{where} \quad T = 2\left(\frac{1}{|A|\alpha} + \beta\right)t$$

*Proof.* For brevity's sake, we omit the conditioning by $E_{\leqslant\tau}$. For every $s = 1, \ldots, T$, define r.v.

$$Y_s \;=\; e_{i,A}^{\tau+s}$$

Observe that $Y_s = f_s(E_{\leqslant\tau+s})$ for a suitable function $f_s$. Lemma 9 implies

$$\mathbf{P}\left(Y_s = 1 \mid E_{\tau+s-1}\right) = \mathbf{P}\left(\deg_{i,A}^{\tau+s} > 0 \;\middle|\; E_{\tau+s-1}\right) \geqslant p = \frac{|A|\alpha}{2 + 2|A|\alpha\beta}$$

By applying Lemma 7 to $E_{\tau+1}, \ldots, E_{\tau+T}$ and $Y_1, \ldots, Y_T$, we get

$$\mathbf{P}\left(\sum_{s=1}^{T} Y_s \;=\; 0\right) \leqslant \mathbf{P}\left(B(T,p) = 0\right) = (1-p)^T \;\leqslant\; \exp(pT) = \exp(-t)$$

□

## C.1 Flooding in Dynamic Graphs

In what follows, we bound the time required to obtain at least $n/2$ informed nodes. This is the *spreading phase.*

**Lemma 13 (Spreading Phase)** *If $\mathcal{G}([n],\{E_t\}_{t\geqslant 0})$ is $(M,\alpha,\beta)$-stationary then, $\forall \tau \geqslant \hat{T}$ with $\hat{T} = O\left(\left(\frac{1}{n\alpha}+\beta\right)^2 \log^2 n\right)$ it holds that*

$$\mathbf{P}\left(|I_\tau| \;<\; \frac{n}{2}\right) \;\leqslant\; \frac{1}{n^2}$$

*Proof.* (Sketch of). Observe that, for any $|A|$, the bound on $T$ for $t = \Theta(\log n)$ in Lemma 11 satisfies

$$T = \Theta\left(\left(\frac{1}{|A|n^2\alpha^2} + \frac{\beta}{n\alpha} + \frac{|A|\beta^2}{n}\right) + \left(\frac{4}{|A|n\alpha} + 3\beta\right)\log n\right) \;=\; O\left(\left(\frac{1}{n\alpha}+\beta\right)^2 \log n\right)$$

From Lemma 11, after every time interval of $T$ epochs, the size of the set of informed nodes at least doubles as far it is smaller than $n/2$, with high probability (say $1-1/n^2$). By a simple application of the Union Bound, we get that, after a $O(\log n)$ number of such time intervals, with high probability the number of informed nodes is at least $n/2$. □

**Lemma 14 (Saturation Phase)** *Let $\mathcal{G}([n],\{E_t\}_{t\geqslant 0})$ be $(M,\alpha,\beta)$-stationary and assume the flooding process is in some epoch $\tau$ such that $|I_\tau| \geqslant n/2$. Then, w.h.p. all nodes get informed within $O\left(\left(\frac{1}{n\alpha}+\beta\right)\log n\right)$ epochs.*

*Proof.* (Sketch of). Assume that we are in some epoch $\tau$ where $|I_\tau| \geqslant n/2$, then by choosing $A = I_\tau$ and $t = \Theta(\log n)$, Lemma 12 implies that, with high probability, every node gets informed within $O\left(\left(\frac{1}{n\alpha}+\beta\right)\log n\right)$ epochs. Then, by the Union Bound, all nodes gets informed after the same number of epochs.

□

**Proof of Theorem 1.** Thanks to Lemma 13, w.h.p., after $O\left(M\left(\frac{1}{n\alpha}+\beta\right)^2 \log^2 n\right)$ steps the set of informed nodes will be at least $n/2$. Then, from Lemma 14, we have that, after further $O\left(M\left(\frac{1}{n\alpha}+\beta\right)\log n\right)$ steps, w.h.p. all nodes will get informed.

□

# D Flooding in Node-MEG: Proof of Theorem 3

We firstly need some notations. For every $x \in S$, define

$$\Gamma(x) = \{y \in S \mid C(x,y) = 1\}$$

In words, $\Gamma(x)$ is the set of states that are at one hop from state $x$. For any node $i \in [n]$, let $\nu_i$ be a probability distribution over the set of states $S$. The symbol $\nu$ (without subscript) will denote the product probability distribution over $\prod_{i\in[n]} S$. Assuming that nodes are random with the probability distribution $\nu$, for every $i,j \in [n]$ and for every $A \in [n]-\{i\}$, define binary random variables $e^\nu_{i,j}$ and $e^\nu_{i,A}$,

$$\begin{aligned} e^\nu_{i,j} &= 1 \quad \text{if nodes } i \text{ and } j \text{ are connected} \\ e^\nu_{i,A} &= 1 \quad \text{if node } i \text{ is connected to some node in } A \end{aligned}$$

It is easy to verify that the followings hold

$$\mathbf{P}\left(e^{\nu}_{i,j}\right) = \sum_{x\in S} \nu_i(x)\nu_j(\Gamma(x))$$
$$\mathbf{P}\left(\overline{e^{\nu}_{i,A}}\right) = \sum_{x\in S} \nu_i(x) \prod_{j\in A}(1-\nu_j(\Gamma(x)))$$

where, for any binary random variable $X$, $\overline{X}$ is the complementary random variable defined as $\overline{X} = 1$ iff $X = 0$. Notice that $\nu_j(\Gamma(x))$ is the probability that node $j$ is connected to a fixed node in state $x$. Let $\pi$ be the stationary probability distribution of the Markov chain $\mathcal{M}$. With an abuse of notation, we denote by $\pi$ also the probability distribution over $\prod_{i\in[n]} S$ given by the product of $\pi$s. For the sake of simplicity, we omit the explicit dependence on $\pi$ of random variables and probabilities. Thus, we write $e_{i,j}$, $e_{i,A}$ to mean, respectively $e^{\pi}_{i,j}$, $e^{\pi}_{i,A}$.

**Lemma 15** *Let* $\mathrm{NM} = NM(n, \mathcal{M}, C)$ *be a node-MEG such that*

$$P_{\mathrm{NM2}} \leqslant \eta \left(P_{\mathrm{NM}}\right)^2 \qquad \textit{for some } \eta \geqslant 1$$

*Then,*

$$\forall i,j \in [n] \forall A \subseteq [n]-\{i,j\} \qquad \mathbf{P}\left(e_{i,A}\cdot e_{j,A}\right) \leqslant 17\eta \mathbf{P}\left(e_{i,A}\right)\mathbf{P}\left(e_{j,A}\right)$$

*Proof.* For the sake of convenience, let $q(x)$ and $q(x,y)$ denote, respectively, $\pi(\Gamma(x))$ and $\pi(\Gamma(x)\cup\Gamma(y))$. Define

$$V = \left\{x \in S \mid q(x) > \frac{1}{\sqrt{|A|}}\right\}$$

**Claim 1** *For every* $k \in [n]-A$*, it holds that*

$$\mathbf{P}\left(e_{k,A}\right) \geqslant \frac{\sqrt{|A|}}{2}\left(P_{\mathrm{NM}} - \sum_{x\in V}\pi(x)q(x)\right) + \frac{1}{2}\pi(V)$$

*Proof.* We upper bound $\mathbf{P}\left(\overline{e_{k,A}}\right)$. Clearly, it holds that

$$\mathbf{P}\left(\overline{e_{k,A}}\right) = \sum_{x\in V}\pi(x)(1-q(x))^{|A|} + \sum_{x\in S-V}\pi(x)(1-q(x))^{|A|} \tag{8}$$

If $x \in V$, then

$$(1-q(x))^{|A|} \leqslant e^{-q(x)|A|} \leqslant e^{-\sqrt{|A|}} \leqslant e^{-1} \tag{9}$$

If $x \in S - V$, then

$$(1-q(x))^{|A|} \leqslant e^{-q(x)|A|} \leqslant 1 - \frac{q(x)|A|}{2\sqrt{|A|}} = 1 - \frac{\sqrt{|A|}}{2}q(x) \tag{10}$$

where we used the inequality $e^{-x} \leqslant 1 - x/(2\alpha)$ (that holds for $\alpha \geqslant 1$ and any $0 \leqslant x \leqslant \alpha$). By combining Inequalities (8), (9), and (10), we obtain

$$\begin{aligned}
\mathbf{P}\left(\overline{e_{k,A}}\right) &\leqslant \frac{1}{e}\sum_{x\in V}\pi(x) + \sum_{x\in S-V}\pi(x)\left(1-\frac{\sqrt{|A|}}{2}q(x)\right) \\
&= \frac{1}{e}\pi(V) + \sum_{x\in S}\pi(x)\left(1-\frac{\sqrt{|A|}}{2}q(x)\right) - \sum_{x\in V}\pi(x)\left(1-\frac{\sqrt{|A|}}{2}q(x)\right) \\
&= \frac{1}{e}\pi(V) + 1 - \frac{\sqrt{|A|}}{2}P_{\mathrm{NM}} - \sum_{x\in V}\pi(x) + \frac{\sqrt{|A|}}{2}\sum_{x\in V}\pi(x)q(x) \\
&= 1 - \left(1-\frac{1}{e}\right)\pi(V) - \frac{\sqrt{|A|}}{2}\left(P_{\mathrm{NM}} - \sum_{x\in V}\pi(x)q(x)\right) \\
&\leqslant 1 - \frac{1}{2}\pi(V) - \frac{\sqrt{|A|}}{2}\left(P_{\mathrm{NM}} - \sum_{x\in V}\pi(x)q(x)\right)
\end{aligned}$$

and the thesis immediately follows. □

We distinguish two cases. First we assume that the following holds

$$\sum_{x\in V}\pi(x)q(x) > \frac{1}{2}P_{\mathrm{NM}} \tag{11}$$

It holds that

$$\begin{aligned}
\frac{1}{\pi(V)}\left(\sum_{x\in V}\pi(x)q(x)\right)^2 &\leqslant \sum_{x\in V}\pi(x)q(x)^2 \qquad \text{(by Jensen's inequality)} \\
&\leqslant \eta\left(P_{\mathrm{NM}}\right)^2 \qquad \text{(by lemma's hypothesis)} \\
&< 4\eta\left(\sum_{x\in V}\pi(x)q(x)\right)^2 \qquad \text{(by Assumption (11))}
\end{aligned}$$

From this, it is immediate to derive that

$$\pi(V) \geqslant \frac{1}{4\eta}$$

Thus, thanks to Claim 1 we obtain

$$\mathbf{P}\left(e_{k,A}\right) \geqslant \frac{\sqrt{|A|}}{2}\left(P_{\mathrm{NM}} - \sum_{x\in V}\pi(x)q(x)\right) + \frac{1}{2}\pi(V) \geqslant \frac{1}{8\eta}$$

It follows that

$$\mathbf{P}\left(e_{i,A}\cdot e_{j,A}\right) \leqslant \mathbf{P}\left(e_{i,A}\right)\frac{8\eta}{8\eta} \leqslant 8\eta\mathbf{P}\left(e_{i,A}\right)\mathbf{P}\left(e_{j,A}\right)$$

Now, we consider the opposite case:

$$\sum_{x \in V} \pi(x) q(x) \leqslant \frac{1}{2} P_{\mathrm{NM}} \tag{12}$$

For any two binary r. v. $X$ and $Y$, it holds that $X \cdot Y = \overline{(\overline{X} \vee \overline{Y})}$ and thus $\mathbf{P}(X \cdot Y) = \mathbf{P}\left(\overline{X} \cdot \overline{Y}\right) + 1 - \mathbf{P}\left(\overline{X}\right) - \mathbf{P}\left(\overline{Y}\right)$. Hence,

$$\mathbf{P}(e_{i,A} \cdot e_{j,A}) = \mathbf{P}(\overline{e_{i,A}} \cdot \overline{e_{j,A}}) + 1 - \mathbf{P}(\overline{e_{i,A}}) - \mathbf{P}(\overline{e_{j,A}}) \tag{13}$$

First, we focus on upper bounding $\mathbf{P}(\overline{e_{i,A}} \cdot \overline{e_{j,A}})$. It holds that

$$\begin{aligned}
\mathbf{P}(\overline{e_{i,A}} \cdot \overline{e_{j,A}}) &= \sum_{x \in S} \sum_{y \in S} \pi(x)\pi(y) \prod_{h \in A} \pi(S - (\Gamma(x) \cup \Gamma(y))) \\
&= \sum_{x,y \in S} \pi(x)\pi(y) \left(1 - \pi(\Gamma(x) \cup \Gamma(y))\right)^{|A|}
\end{aligned}$$

Define

$$R = \{(x,y) \in S \times S \mid \Gamma(x) \cap \Gamma(y) \neq \emptyset\} \quad \text{and} \quad \overline{R} = S \times S - R$$

Observe that if $(x,y) \in \overline{R}$ then $q(x,y) = q(x) + q(y)$. Thus, it holds that

$$\begin{aligned}
\mathbf{P}(\overline{e_{i,A}} \cdot \overline{e_{j,A}}) &= \sum_{(x,y) \in \overline{R}} \pi(x)\pi(y)(1 - q(x,y))^{|A|} + \sum_{(x,y) \in R} \pi(x)\pi(y)(1 - q(x,y))^{|A|} \\
&= \sum_{(x,y) \in \overline{R}} \pi(x)\pi(y)(1 - q(x) - q(y))^{|A|} + \sum_{(x,y) \in R} \pi(x)\pi(y)(1 - q(x,y))^{|A|} \\
&= \Lambda_1 + \Lambda_2
\end{aligned} \tag{14}$$

where

$$\begin{aligned}
\Lambda_1 &= \sum_{x,y \in S} \pi(x)\pi(y)(\max\{1 - q(x) - q(y), 0\})^{|A|} \\
\Lambda_2 &= \sum_{(x,y) \in R} \pi(x)\pi(y) \left[(1 - q(x,y))^{|A|} - (\max\{1 - q(x) - q(y), 0\})^{|A|}\right]
\end{aligned}$$

To bound $\Lambda_1$ we use the inequality $\max\{1 - a - b, 0\} \leqslant (1-a)(1-b)$, that holds for any $a$ and $b$ with $0 \leqslant a, b \leqslant 1$:

$$\begin{aligned}
\Lambda_1 &\leqslant \sum_{x,y \in S} \pi(x)\pi(y)(1 - q(x))^{|A|}(1 - q(y))^{|A|} \\
&= \sum_{x \in S} \pi(x)(1 - q(x))^{|A|} \sum_{y \in S} \pi(y)(1 - q(y))^{|A|} \\
&= \mathbf{P}(\overline{e_{i,A}})\, \mathbf{P}(\overline{e_{j,A}})
\end{aligned} \tag{15}$$

By combining Inequalities (13), (14), and (15), we obtain

$$\begin{aligned}\mathbf{P}\left(e_{i,A}\cdot e_{j,A}\right) &\leqslant 1-\mathbf{P}\left(\overline{e_{i,A}}\right)-\mathbf{P}\left(\overline{e_{j,A}}\right)+\mathbf{P}\left(\overline{e_{i,A}}\right)\mathbf{P}\left(\overline{e_{j,A}}\right)+\Lambda_2\\ &=\left(1-\mathbf{P}\left(\overline{e_{i,A}}\right)\right)\left(1-\mathbf{P}\left(\overline{e_{j,A}}\right)\right)+\Lambda_2\\ &=\mathbf{P}\left(e_{i,A}\right)\mathbf{P}\left(e_{j,A}\right)+\Lambda_2 \qquad (16)\end{aligned}$$

To upper bound $\Lambda_2$, we use the following

**Claim 2** *If* $0\leqslant b\leqslant 1$ *and* $1\leqslant 1-a+b\leqslant 1$, *then, for any integer* $k\geqslant 1$,

$$(1-a+b)^k-(\max\{1-a,0\})^k\leqslant kb$$

*Proof.* By distinguishing the two cases $1-a\leqslant 0$ and $1-a>0$, and in the latter by induction on $k$. □

Let $\hat{q}(x,y)$ denote $\pi(\Gamma(x)\cap\Gamma(y))$. Observe that

$$q(x,y)=q(x)+q(y)-\hat{q}(x,y)$$

From Claim 2 with $a=q(x)+q(y)$ and $b=\hat{q}(x,y)$, we get

$$(1-q(x,y))^{|A|}-(\max\{1-q(x)-q(y),0\})^{|A|}\leqslant |A|\hat{q}(x,y)$$

It follows that

$$\begin{aligned}\Lambda_2 &\leqslant |A|\sum_{(x,y)\in R}\pi(x)\pi(y)\hat{q}(x,y)\\ &=|A|\sum_{x,y\in S}\pi(x)\pi(y)\hat{q}(x,y) \qquad (\text{since } (x,y)\notin R\Rightarrow \hat{q}(x,y)=0))\\ &=|A|\sum_{x,y\in S}\pi(x)\pi(y)\sum_{z\in S}\pi(z)C(z,x)C(z,y)\\ &=|A|\sum_{z\in S}\pi(z)\sum_{x\in S}\pi(x)C(z,x)\sum_{y\in S}\pi(y)C(z,y)\\ &=|A|\sum_{z\in S}\pi(z)q(z)^2\\ &\leqslant \eta|A|\left(P_{\mathrm{NM}}\right)^2 \qquad (\text{from lemma's hypothesis})\end{aligned}$$

By combining this with Inequality (16), we obtain

$$\mathbf{P}\left(e_{i,A}\cdot e_{j,A}\right)\leqslant \mathbf{P}\left(e_{i,A}\right)\mathbf{P}\left(e_{j,A}\right)+\eta|A|\left(P_{\mathrm{NM}}\right)^2 \qquad (17)$$

From Claim 1 and Hypothesis (12) we get

$$\mathbf{P}\left(e_{k,A}\right)\geqslant\frac{\sqrt{|A|}}{2}\left(P_{\mathrm{NM}}-\sum_{x\in V}\pi(x)q(x)\right)\geqslant\frac{\sqrt{|A|}}{2}\left(P_{\mathrm{NM}}-\frac{1}{2}P_{\mathrm{NM}}\right)=\frac{\sqrt{|A|}}{4}P_{\mathrm{NM}}$$

In conclusion, from this and Inequality (17) we get the thesis:

$$\begin{aligned}\mathbf{P}\left(e_{i,A}\cdot e_{j,A}\right) &\leqslant \mathbf{P}\left(e_{i,A}\right)\mathbf{P}\left(e_{j,A}\right)+\eta|A|\left(P_{\mathrm{NM}}\right)^2\\ &\leqslant \mathbf{P}\left(e_{i,A}\right)\mathbf{P}\left(e_{j,A}\right)+\eta|A|\frac{4\mathbf{P}\left(e_{i,A}\right)}{\sqrt{|A|}}\frac{4\mathbf{P}\left(e_{j,A}\right)}{\sqrt{|A|}}\\ &= \mathbf{P}\left(e_{i,A}\right)\mathbf{P}\left(e_{j,A}\right)+16\eta\mathbf{P}\left(e_{i,A}\right)\mathbf{P}\left(e_{j,A}\right)\leqslant 17\eta\mathbf{P}\left(e_{i,A}\right)\mathbf{P}\left(e_{j,A}\right)\end{aligned}$$

□

**Lemma 16** *For every $i=1,\dots,k$, let $\psi_i$ and $\zeta_i$ be any two probability distributions over any domain $\Omega_i$. Let $\psi$ and $\zeta$ denote the product probability distributions over $\prod_{i=1}^{k}\Omega_i$ of $\psi_i$s and $\zeta_i$s, respectively. Then, it holds that*

$$||\psi-\zeta||_{\mathrm{TV}}\leqslant\sum_{i=1}^{k}||\psi_i-\zeta_i||_{\mathrm{TV}}$$

*Proof.* Let $\Omega=\prod_{i=1}^{k}\Omega_i$. We denote $(x_1,\dots,x_k)$ by $\overline{x}$. For every $i$, let $\Omega^{-i}=\prod_{j=1,j\neq i}^{k}\Omega_j$ and let $\overline{x}^{-i}$ denote $(x_1,\dots x_{i-1},x_{i+1},\dots,x_k)$. It holds that

$$\begin{aligned}||\psi-\zeta||_{\mathrm{TV}} &= \frac{1}{2}\sum_{\overline{x}\in\Omega}|\psi(\overline{x})-\zeta(\overline{x})|\\ &= \frac{1}{2}\sum_{\overline{x}\in\Omega}\left|\prod_{j=1}^{k}\psi_j(x_j)-\prod_{j=1}^{k}\zeta_j(x_j)\right|\\ &= \frac{1}{2}\sum_{\overline{x}\in\Omega}\left|\prod_{j=1}^{k}\psi_j(x_j)+\sum_{i=2}^{k}\prod_{j=1}^{i-1}\zeta_j(x_j)\prod_{j=i}^{k}\psi_j(x_j)-\prod_{j=1}^{k}\zeta_j(x_j)-\sum_{i=2}^{k}\prod_{j=1}^{i-1}\zeta_j(x_j)\prod_{j=i}^{k}\psi_j(x_j)\right|\\ &= \frac{1}{2}\sum_{\overline{x}\in\Omega}\left|\sum_{i=1}^{k}\prod_{j=1}^{i-1}\zeta_j(x_j)\prod_{j=i}^{k}\psi_j(x_j)-\sum_{i=1}^{k}\prod_{j=1}^{i}\zeta_j(x_j)\prod_{j=i+1}^{k}\psi_j(x_j)\right|\\ &= \frac{1}{2}\sum_{\overline{x}\in\Omega}\left|\sum_{i=1}^{k}\psi_i(x_i)\prod_{j=1}^{i-1}\zeta_j(x_j)\prod_{j=i+1}^{k}\psi_j(x_j)-\sum_{i=1}^{k}\zeta_i(x_i)\prod_{j=1}^{i-1}\zeta_j(x_j)\prod_{j=i+1}^{k}\psi_j(x_j)\right|\\ &= \frac{1}{2}\sum_{\overline{x}\in\Omega}\left|\sum_{i=1}^{k}(\psi_i(x_i)-\zeta_i(x_i))\prod_{j=1}^{i-1}\zeta_j(x_j)\prod_{j=i+1}^{k}\psi_j(x_j)\right|\\ &\leqslant \frac{1}{2}\sum_{\overline{x}\in\Omega}\sum_{i=1}^{k}|\psi_i(x_i)-\zeta_i(x_i)|\prod_{j=1}^{i-1}\zeta_j(x_j)\prod_{j=i+1}^{k}\psi_j(x_j)\\ &= \frac{1}{2}\sum_{i=1}^{k}\sum_{x_i\in\Omega_i}|\psi_i(x_i)-\zeta_i(x_i)|\sum_{\overline{x}^{-i}\in\Omega^{-i}}\prod_{j=1}^{i-1}\zeta_j(x_j)\prod_{j=i+1}^{k}\psi_j(x_j) \qquad (18)\end{aligned}$$

Observe that

$$\sum_{\overline{x}^{-i}\in\Omega^{-i}} \prod_{j=1}^{i-1} \zeta_j(x_j) \prod_{j=i+1}^{k} \psi_j(x_j) = \sum_{(x_1,\dots x_{i-1})\in\prod_{j=1}^{i-1}\Omega_j} \prod_{j=1}^{i-1} \zeta_j(x_j) \sum_{(x_{i+1},\dots x_k)\in\prod_{j=i+1}^{k}\Omega_j} \prod_{j=i+1}^{k} \psi_j(x_j)$$

$$= \sum_{(x_1,\dots x_{i-1})\in\prod_{j=1}^{i-1}\Omega_j} \prod_{j=1}^{i-1} \zeta_j(x_j)$$

$$= 1$$

Hence, from this and Inequality (18) the thesis follows. □

**Lemma 17** *Let* NM $= NM(n, \mathcal{M}, C)$ *be a node-MEG such that*

$$P_{\mathrm{NM2}} \leqslant \eta \left(P_{\mathrm{NM}}\right)^2 \qquad \textit{for some } \eta \geqslant 1 \tag{19}$$

*Moreover, let* $\nu$ *be a product probability distribution such that*

$$\forall i \in [n] \qquad ||\nu_i - \pi||_{\mathrm{TV}} \leqslant \frac{\left(P_{\mathrm{NM}}\right)^2}{2n} \tag{20}$$

*Then, for any* $i, j \in [n]$ *and for any* $A \subseteq [n] - \{i, j\}$,

*(a)* $\mathbf{P}\left(e^{\nu}_{i,j}\right) \geqslant \frac{P_{\mathrm{NM}}}{2}$

*(b)* $\mathbf{P}\left(e^{\nu}_{i,A} \cdot e^{\nu}_{j,A}\right) \leqslant 72\eta \mathbf{P}\left(e^{\nu}_{i,A}\right) \mathbf{P}\left(e^{\nu}_{j,A}\right)$

*Proof.* Let $C_{i,j} = \{(x_1, \dots, x_n) \in \prod_{i=1}^{n} S \mid C(x_i, x_j) = 1\}$. It holds that

$$\begin{aligned} \mathbf{P}\left(e^{\nu}_{i,j}\right) &= \nu(C_{i,j}) \\ &\geqslant \pi(C_{i,j}) - n\frac{\left(P_{\mathrm{NM}}\right)^2}{2n} \qquad \text{(from (20) and Lemma 16)} \\ &= P_{\mathrm{NM}} - \frac{\left(P_{\mathrm{NM}}\right)^2}{2} \geqslant \frac{P_{\mathrm{NM}}}{2} \end{aligned}$$

This proves (a).

For any $k \in [n]$ and for any $A \subseteq [n] - \{k\}$, let $C_{k,A} = \{(x_1, \dots, x_n) \in \prod_{i=1}^{n} S \mid \exists h \in A : C(x_k, x_h) = 1\}$. For any $i, j \in [n]$ and for any $A \subseteq [n] - \{i, j\}$, let $C_{i,j,A} = \{(x_1, \dots, x_n) \in \prod_{i=1}^{n} S \mid \exists h, k \in A : C(x_i, x_h) = 1 \wedge C(x_j, x_k) = 1\}$. It holds that

$$\begin{aligned} \mathbf{P}\left(e^{\nu}_{i,A} \cdot e^{\nu}_{j,A}\right) &= \nu(C_{i,j,A}) \\ &\leqslant \pi(C_{i,j,A}) + n\frac{\left(P_{\mathrm{NM}}\right)^2}{2n} \qquad \text{(from (20) and Lemma 16)} \\ &= \mathbf{P}\left(e_{i,A} \cdot e_{j,A}\right) + \frac{\left(P_{\mathrm{NM}}\right)^2}{2} \end{aligned} \tag{21}$$

Since Hypothesis (19) holds, Lemma 15 ensures that

$$\mathbf{P}\left(e_{i,A} \cdot e_{j,A}\right) \leqslant 17\eta \mathbf{P}\left(e_{i,A}\right) \mathbf{P}\left(e_{j,A}\right)$$

From this and Inequality (21) we obtain

$$\begin{aligned}
\mathbf{P}\left(e^{\nu}_{i,A} \cdot e^{\nu}_{j,A}\right) &\leqslant 17\eta \mathbf{P}\left(e_{i,A}\right) \mathbf{P}\left(e_{j,A}\right) + \frac{(P_{\mathrm{NM}})^2}{2} \\
&\leqslant 18\eta \mathbf{P}\left(e_{i,A}\right) \mathbf{P}\left(e_{j,A}\right) \qquad (\text{since } P_{\mathrm{NM}} \leqslant \mathbf{P}\left(e_{i,A}\right), \mathbf{P}\left(e_{j,A}\right)) \\
&= 18\eta \pi(C_{i,A}) \pi(C_{j,A}) \\
&\leqslant 18\eta(\nu(C_{i,A}) + \frac{(P_{\mathrm{NM}})^2}{2})(\nu(C_{j,A}) + \frac{(P_{\mathrm{NM}})^2}{2}) \qquad (\text{from (20) and Lemma 16}) \\
&\leqslant 18\eta(\mathbf{P}\left(e^{\nu}_{i,A}\right) + \frac{P_{\mathrm{NM}}}{2})(\mathbf{P}\left(e^{\nu}_{j,A}\right) + \frac{P_{\mathrm{NM}}}{2}) \qquad (22)
\end{aligned}$$

Since $P_{\mathrm{NM}} = \mathbf{P}\left(e_{i,j}\right) = \pi(\{(x_1, \ldots, x_n) \in \prod_{i=1}^n S \mid C(x_i, x_j) = 1\})$, from (20) and Lemma 16 it derives that $P_{\mathrm{NM}} \leqslant \mathbf{P}\left(e^{\nu}_{i,j}\right) + ((P_{\mathrm{NM}})^2)/2$. It follows that $P_{\mathrm{NM}}/2 \leqslant P_{\mathrm{NM}} - ((P_{\mathrm{NM}})^2)/2 \leqslant \mathbf{P}\left(e^{\nu}_{i,j}\right) \leqslant \mathbf{P}\left(e^{\nu}_{i,A}\right)$. From this and Inequality (22) we get

$$\mathbf{P}\left(e^{\nu}_{i,A} \cdot e^{\nu}_{j,A}\right) \leqslant 18\eta(2\mathbf{P}\left(e^{\nu}_{i,A}\right))(2\mathbf{P}\left(e^{\nu}_{j,A}\right)) = 72\eta \mathbf{P}\left(e^{\nu}_{i,A}\right) \mathbf{P}\left(e^{\nu}_{j,A}\right)$$

□

**Proof of Theorem 3.**
Let $\mathcal{G} = \mathcal{G}([n], \{E_t\}_{t \geqslant 0})$ be the dynamic graph yielded by $NM(n, \mathcal{M}, C)$ and let $e^t_{i,j}$ and $e^t_{i,A}$ be the relative random variables. Consider $M = T_{\mathrm{mix}} \log(2n/(P_{\mathrm{NM}})^2)$ be the duration of an epoch. From standard results on Markov chains, it derives that, for any node $i$, whatever the probability distribution at time $t$ the probability distribution $\nu_i$ at time $t + M$ is such that

$$||\nu_i - \pi||_{\mathrm{TV}} \leqslant 2^{-\log(2n/(P_{\mathrm{NM}})^2)} = \frac{(P_{\mathrm{NM}})^2}{2n} \qquad (23)$$

This and the theorem's hypothesis satisfy the hypotheses of Lemma 17 and in turn it implies that

$$\begin{aligned}
&\mathbf{P}\left(e^{\tau M}_{i,j} \mid E_{t \leqslant (\tau-1)M}\right) \geqslant \frac{P_{\mathrm{NM}}}{2} &(24)\\
&\mathbf{P}\left(e^{\tau M}_{i,A} \cdot e^{\tau M}_{j,A} \mid E_{t \leqslant (\tau-1)M}\right) \leqslant 72\eta \mathbf{P}\left(e^{\tau M}_{i,A} \mid E_{t \leqslant (\tau-1)M}\right) \mathbf{P}\left(e^{\tau M}_{j,A} \mid E_{t \leqslant (\tau-1)M}\right) &(25)
\end{aligned}$$

Hence, Inequalities (25) and (24) show that $\mathcal{G}$ is a $(M, P_{\mathrm{NM}}/2, 72\eta)$-stationary dynamic graph. Then, from Theorem 1, the flooding time in $\mathcal{G}$ is, with high probability,

$$\mathrm{O}\left(T_{\mathrm{mix}} \log(2n/(P_{\mathrm{NM}})^2) \left(\frac{2}{nP_{\mathrm{NM}}} + 72\eta\right)^2 \log^2 n\right)$$

and, taking into account the hypothesis $P_{\mathrm{NM}} \geqslant 1/n^{\mathrm{O}(1)}$, it is

$$\mathrm{O}\left(T_{\mathrm{mix}} \left(\frac{1}{nP_{\mathrm{NM}}} + \eta\right)^2 \log^3 n\right)$$

□

### D.1 Flooding in Classic Mobility Models: Proofs

**Proof of Corollary 4.**
Let $\pi$ be the stationary distribution of the Markov chain $\mathcal{M}$ of NM. For any $x \in S$, let $u(x)$ be the point in $\mathcal{R}$ where a node is when its state is $x$. By assuming that the states of the nodes are random with $\pi$, let $q(x)$ be the probability that a fixed node is connected to another fixed node being in state $x$. Clearly,

$$q(x) = \sum_{y \in S \,:\, u(y) \in D(u(x),r)} \pi(y)$$

Since NM is a sufficiently refined discrete version of the random trip model $\mathcal{T}$, it holds that

$$q(x) \approx \int_{D(u(x),r)} F_{\mathcal{T}}(u)du \tag{26}$$

From Hypothesis (b), for every $v \in \mathcal{B}_r$,

$$\forall u \in D(v,r) \qquad F_{\mathcal{T}}(u) \geqslant \frac{1}{\delta \mathrm{vol}(\mathcal{R})}$$

This implies that

$$\int_{D(v,r)} F_{\mathcal{T}}(u)du \geqslant \frac{\mathrm{vol}(D(v,r))}{\delta \mathrm{vol}(\mathcal{R})} = \frac{c_d r^d}{\delta \mathrm{vol}(\mathcal{R})} \tag{27}$$

where $c_d$ is a constant depending only on $d$. By combining (26) and (27), we have that, for every $x \in S$ with $u(x) \in \mathcal{B}_r$,

$$q(x) \gtrsim \frac{c_d r^d}{\delta \mathrm{vol}(\mathcal{R})} \tag{28}$$

It holds that

$$\begin{aligned}
P_{\mathrm{NM}} &= \sum_{x \in S} \pi(x) q(x) \\
&\geqslant \sum_{x \in S \,:\, u(x) \in \mathcal{B}_r} \pi(x) q(x) \\
&\gtrsim \frac{c_d r^d}{\delta \mathrm{vol}(\mathcal{R})} \sum_{x \in S \,:\, u(x) \in \mathcal{B}_r} \pi(x) \qquad \text{(from Inequality (28))} \\
&\gtrsim \frac{c_d r^d}{\delta \mathrm{vol}(\mathcal{R})} \int_{\mathcal{B}_r} F_{\mathcal{T}}(u)du \\
&\gtrsim \frac{c_d r^d}{\delta \mathrm{vol}(\mathcal{R})} \frac{\mathrm{vol}(\mathcal{B}_r)}{\delta \mathrm{vol}(\mathcal{R})} \qquad \text{(from Hypothesis (b))} \\
&\gtrsim \frac{\lambda c_d r^d}{\delta^2 \mathrm{vol}(\mathcal{R})} \qquad \text{(from Hypothesis (b))}
\end{aligned} \tag{29}$$

From (26) and Hypothesis (a), it derives that, for every $x \in S$,

$$q(x) \approx \int_{D(u(x),r)} F_{\mathcal{T}}(u)du \lesssim \frac{\delta \mathrm{vol}(D(u(x),r))}{\mathrm{vol}(\mathcal{R})} = \frac{\delta c_d r^d}{\mathrm{vol}(\mathcal{R})} \tag{30}$$

It holds that

$$\begin{aligned} P_{\mathrm{NM2}} &= \sum_{x \in S} \pi(x) q(x)^2 \\ &\lesssim \left( \frac{\delta c_d r^d}{\mathrm{vol}(\mathcal{R})} \right)^2 \sum_{x \in S} \pi(x) \qquad \text{(from Inequality (30))} \\ &= \frac{\delta^2 c_d^2 r^{2d}}{\mathrm{vol}(\mathcal{R})^2} \end{aligned}$$

From this and Inequality (29), we get

$$P_{\mathrm{NM2}} \lesssim \frac{\delta^2 c_d^2 r^{2d}}{\mathrm{vol}(\mathcal{R})^2} = \frac{\delta^6}{\lambda^2} \left( \frac{\lambda c_d r^d}{\delta^2 \mathrm{vol}(\mathcal{R})} \right)^2 \lesssim \frac{\delta^6}{\lambda^2} \left( P_{\mathrm{NM}} \right)^2$$

It follows that the node-MEG NM satisfies the hypotheses of Theorem 3 with $\eta = \delta^6/\lambda^2$ and thus, with high probability, the flooding time is

$$\mathrm{O}\left( T_{\mathrm{mix}} \left( \frac{1}{n P_{\mathrm{NM}}} + \frac{\delta^6}{\lambda^2} \right)^2 \log^3 n \right) \leqslant \mathrm{O}\left( T_{\mathrm{mix}} \left( \frac{\delta^2 \mathrm{vol}(R)}{\lambda n r^d} + \frac{\delta^6}{\lambda^2} \right)^2 \log^3 n \right)$$

where we have used Inequality (29). □

**Proof of Corollary 5.**

It is easy to see that, any node-MEG yielded by a random-path model is a Markov Trace Model (MTM), a general class of models introduced in [14]. Since RP is simple and reversible, Theorem 11 in [14] implies that the stationary distribution $\pi$ of $\mathcal{M}$ is uniform. For any state $x \in S$, let $u(x) \in V$ be the point where a node is when its state is $x$. By assuming that the states of the nodes are random with $\pi$, let $q(x)$ be the probability that a fixed node is connected to another fixed node being in state $x$. Since $\pi$ is uniform and RP is simple, it holds that

$$q(x) = \sum_{y \in S \,:\, C(y,x)=1} \pi(x) = \frac{1}{|S|} |\{y \in S \mid u(y) = u(x)\}| = \frac{\#_{\mathcal{P}}(u(x))}{|S|} \tag{31}$$

It follows that

$$P_{\mathrm{NM}} = \sum_{x \in S} \pi(x) q(x) = \frac{1}{|S|^2} \sum_{x \in S} \#_{\mathcal{P}}(u(x)) = \frac{1}{|S|^2} \sum_{u \in V} \#_{\mathcal{P}}(u)^2 \tag{32}$$

where the last equality derives from $|\{y \in S \mid u(y) = u\}| = \#_{\mathcal{P}}(u)$. Thanks to Jensen's inequality it holds that

$$\frac{\sum_{u \in V} \#_{\mathcal{P}}(u)^2}{|V|} \geqslant \left( \frac{\sum_{u \in V} \#_{\mathcal{P}}(u)}{|V|} \right)^2 = \frac{|S|^2}{|V|^2}$$

Thus, from (32), we have

$$P_{\rm NM} \geqslant \frac{1}{|V|} \tag{33}$$

From (31), it holds that

$$P_{\rm NM2} = \sum_{x \in S} \pi(x) q(x)^2 = \frac{1}{|S|^3} \sum_{x \in S} \#_{\mathcal{P}}(u(x))^2 = \frac{1}{|S|^3} \sum_{u \in V} \#_{\mathcal{P}}(u)^3$$

Since $RP$ is $\delta$-regular and (33) holds, it follows that

$$P_{\rm NM2} = \frac{1}{|S|^3} \sum_{u \in V} \#_{\mathcal{P}}(u)^3 \leqslant \frac{1}{|S|^3} \sum_{u \in V} \left( \delta \frac{\sum_{v \in V} \#_{\mathcal{P}}(v)}{|V|} \right)^3 = \frac{\delta^3}{|V|^2} \leqslant \delta^3 \left(P_{\rm NM}\right)^2$$

From this and taking into account the hypothesis $|V| \leqslant n^{\mathrm{O}(1)}$ and (33), Theorem 3 can be applied with $\eta = \delta^3$ obtaining that, with high probability, the flooding time is

$$\mathrm{O}\left( T_{\rm mix} \left( \frac{|V|}{n} + \delta^3 \right)^2 \log^3 n \right)$$

□